\documentclass[12pt]{article}
\usepackage{epsf}
\begin{document}
\newcommand {\sheptitle}
{Bimaximal Mixings from the Texture of the Right-handed Majorana Neutrino Mass
 Matrix}
\newcommand {\shepauthor}
{N.Nimai Singh$^{\dag,}$\footnote{Corresponding author.
\\ {\it {E-mail address}}: nimai@guphys.cjb.net} and Mahadev Patgiri$^{\ddag}$}
\newcommand{\shepaddress}
{$^{\dag}$ Department of Physics, Gauhati University, Guwahati-781014, India\\
$^{\ddag}$ Department of Physics, Cotton College, Guwahati-781001, India}
\newcommand{\shepabstract}
{We study the origin of neutrino masses and mixing angles which can accomodate
 the  LMA MSW solutions  of the solar neutrino
anomaly  as well as the solution of the atmospheric neutrino problem, within
 the framework of the see-saw mechanism. We employ the diagonal form of the 
Dirac neutrino mass matrices with the physical masses as diagonal elements in
 the hierarchical order. Such choice has been
motivated from the fact that the known CKM angles for the quark sector, are
 relatively small. We consider both possibilities where the  Dirac neutrino
 mass  matrix is
either the charged lepton or the up-quark mass matrix within the framework of 
SO(10) GUT with or without supersymmetry. The non-zero texture of
the right-handed Majorana  neutrino mass matrix $M_{R}$ is used  for
 the generation of the desired
bimaximal mixings in a model independent way. Both hierarchical and
 inverted hierarchical models of the left-handed Majorana neutrino mass
 matrices are generated and then discussed with examples.}
%88888888888888888888888888888888888888888888888888888888888888888
\begin{titlepage}
\begin{flushright}
hep-ph/0111319
\end{flushright}
\begin{center}
{\large{\bf\sheptitle}}
\bigskip \\
\shepauthor
\\
\mbox{}
\\
{\it \shepaddress}
\\
\vspace{.5in}
{\bf Abstract} \bigskip \end{center}\setcounter{page}{0}
\shepabstract
\end{titlepage}
\section{Introduction}
The latest results from the Super-Kamiokande experiments [1,2] on atmospheric
 and solar neutrino oscillations, indicate strong
evidence for the existence of the non-zero neutrino masses and mixing angles.
 The mass-squared difference required by the solution of the
atmospheric neutrino  anomaly [2] is found out  to be $\Delta m_{23}^{2}=
|m_{3}^{2}-m_{2}^{2}|\sim (1.5-5)\times 10^{-3} eV^{2}$
at 90\% CL    and the best-fit value is centred around $\Delta m_{23}^{2}
\approx 3\times10^{-2} eV^{2}$. The lower bound on
the maximal mixing is $sin^{2}2\theta_{23}\geq 0.88$. In case of the solar
 neutrino anomaly the exact solution is still not clear.
The main solutions to the solar neutrino problem [1] are the large mixing
angle (LMA) MSW, small mixing
 angle(SMA) MSW, the low mass splitting (LOW) solution, and
the vacuum(VAC) solution as shown in Table-I.\\
Table-I: Latest analysis of the best fits on solar neutrino data [3,4].
\begin{center}
\begin{tabular}{lll} \hline \hline
\it{Type} &  \it{$\Delta m_{12}^{2} (eV^{2})$} &  \it{$ sin^{2}2\theta_{12}$}\\ \hline
LMA  & $4.2\times 10^{-5}$  & $0.8163$\\
SMA  & $5.1\times 10^{-6}$  & $0.0027$\\
LOW  & $9.6\times 10^{-8}$  & $0.93$\\
VAC  & $8.0\times 10^{-10}$ & $0.93$\\ \hline
\end{tabular}
\end{center}
The Super-Kamiokande results disfavour the SMA MSW solution, VAC solution,
 and the oscillation of
$\nu_{e}$ into a sterile neutrino. Thus the preferred solution at present
 seems to be the LMA MSW solution although a                        similar
 solution with LOW solution is possible. A typical latest best-fit value for
 LMA solution is found to be
$\Delta m_{21}^{2}=|m_{2}^{2}-m_{1}^{2}| =4.2 \times 10^{-5} eV^{2}$[4] with
 the corresponding mixing angle [5] $sin^{2}2\theta_{12}=0.8163$ but  the upper
 limit [5] is set at  $sin^{2}2\theta_{12} \leq 0.988$. With this LMA MSW solution the splitting parameter defined by
 $\xi=\Delta m_{12}^{2}/\Delta m_{23}^{2}$ is obtained
 as $\xi=1.4\times10^{-2}$ [6]. The CHOOZ experiment [7] limits
 $sin^{2}2\theta_{13}$ to the range from 0.1 to 0.3 over the Super-Kamiokande
 preferred range of $\Delta m_{23}^{2}$. We exclude the LSND result [8] 
in the present analysis and confine to three neutrino species only.

The available observational values on $\Delta m_{12}^{2}$ and
 $\Delta m_{23}^{2}$ are not sufficient to conclude whether the neutrino
masses are hierarchical similar to the known quark and charge lepton masses,
 or inversely hierarchical [3]. Instead, both cases are possible.
 In the heirarchical model the splitting parameter
 $\xi=\Delta m_{12}^{2}/\Delta m_{23}^{2}$ is simply $\xi=m_{\nu2}^{2}/m_{\nu3}^{2}$  where
$m_{\nu3}>>m_{\nu2}>>m_{\nu1}$; but $\xi=2(m_{\nu1}-m_{\nu2})/m_{\nu2}$
 for inverted hierarchical model where $m_{\nu3}<<m_{\nu2}\approx m_{\nu1}$.
We employ the see-saw mechanism [9] for generating the small left-handed
 Majorana neutrino masses $(m_{LL})$ but such prediction has
inherent ambiguities due to the unknown structure of the heavy right-handed 
Majorana mass
 matrix $M_{R}$ which has generally no direct relationship to those of the
 Dirac quark and lepton mass matrices.

We specify our focus on the LMA MSW solution. There are several possible
 solutions which have been suggested in the literature [6,10,11,12,13]
and the most of these attempts employ the non-diagonal texture of the Dirac
 neutrino mass matrix $(m_{LR})$ and the charged lepton mass matrix
$(m_{e})$. It will be interesting to study the non-zero texture of the
 right-handed Majorana mass matrix $M_{R}$ in a  model independent
way while keeping the Dirac neutrino and charged lepton mass matrices in
 the diagonal form. The left-handed Majorana neutrino mass matrix
$(m_{LL})$ thus obtained from the see-saw formula, can explain both LMA MSW
 solar and atmospheric solutions. Both hierarchical and
inverted hierarchical schemes can be incorporated under a single framework.

The paper is organised as follows. In Section 2, we present a general form
 of $M_{R}$ and then generate the left-handed Majorana
neutrino mass matrix, keeping the Dirac neutrino and the charged lepton mass
 matrices in the diagonal form. In Section 3, we present the
bimaximal mixings  for both hierarchical and inverted hierarchical models.
 We conclude with a discussion in Section 4.

\section{Neutrino mass matrix from a non-zero texture of $M_{R}$}
The left-handed Majorana neutrino mass matrix is given by the see-saw
 formula [9]
\begin{equation}
m_{LL}= m_{LR} M_{R}^{-1} m_{LR}^{T}
\end{equation}
where $m_{LR}$ is the Dirac neutrino mass matrix in the left-right (LR)
 convention. The lepton mixing matrix known as
 the MNS mixing matrix [14] is defined as
\begin{equation}
 V_{MNS}=V_{eL} V_{\nu L}^{\dag}
\end{equation}
 where $V_{eL}$ and $V_{\nu L}$ are obtained from the diagonalisation
 of the charged lepton and left-handed Majorana mass matrices,
$$m_{e}^{diag}=V_{eL} m_{e} V_{eR}^{\dag}$$
\begin{equation}
m_{LL}^{diag}= V_{\nu L} m_{LL} V_{\nu L}^{\dag}
\end{equation}
The neutrino flavour eigenstate $\nu_{f}$ is related to the mass 
eigenstate $\nu_{i}$ by the relation $$\nu_{f}=V_{fi}\nu_{i}$$ where $f=e,\mu,\tau$ and $i=1,2,3$. The mixing matrix $V_{fi}$ is now defined 
 by the MNS matrix (2).
From the unitary conditions of the MNS matrix elements and its
 parametrisation by a sequence of three rotations about the 1, 2, 3 axes,
the mixing angles are usually expressed in terms of the elements of
 $V_{MNS}$ as
$$sin^{2}2\theta_{12}=
 \frac{4V^{2}_{e2} V^{2}_{e1}}{(V^{2}_{e2}+V^{2}_{e1})^2},$$
\begin{equation}
sin^{2}2\theta_{23}= \frac{4V^{2}_{\mu 3} V^{2}_{\tau 3}}{(V^{2}_{\mu 3}+V^{2}_{\tau 3})^2}
\end{equation}
In the basis where the charged lepton mass matrix is diagonal, the MNS mixing
 matrix in Eq.(2) reduces to $V_{MNS}= V_{\nu L}^{\dag}$,
and $m_{LL}$ in Eq.(1) is replaced by [15]
\begin{equation}
m_{LL}^{'}= V_{eL} m_{LL} V_{eL}^{\dag}
\end{equation}
where $m_{LR}$ is redefined as $(V_{eL} m_{LR})$ in the see-saw formula in
 Eq.(1).
In order to generate the lepton mixings from the texture of $M_{R}$ only,
 we consider the diagonal form of $m_{e}$ and $m_{LR}$.
This leads to $m_{LL}^{'}= m_{LL}$, and $V_{MNS}= V_{\nu L}^{\dag}$. The origin
 of $M_{R}$ is quite different from those of the Dirac mass
matrices $m_{LR}$ in the underlying grand unified model. Since the Dirac
 neutrino mass matrices are hierarchical in nature and the
CKM mixing angles of the quark sector are relatively small,
 our choices of the diagonal Dirac
 neutrino mass matrix and charged lepton mass matrix are
partly justified. The large mixings for the solar and the atmospheric
 neutrino oscillations will now have their origin from the texture of
$M_{R}$. We parametrise in a model independant way, the most general
 form of $M_{R}$ as
\begin{equation}
M_{R}=
\left(\begin{array}{ccc}
M_{11} & M_{12} & M_{13}\\ M_{21} & M_{22} & M_{23}\\ M_{31} & M_{32} & M_{33}
\end{array}\right)\\
=\left(\begin{array}{ccc}\epsilon_{1}&\sigma &\rho\\
\sigma & \epsilon_{2} & \mu\\ \rho &\mu & \epsilon_{3}
\end{array}\right) v_{R}
\end{equation}
where $v_{R}$ is the vacuum expectation value (VEV) of the Higgs field which
 gives mass to the right-handed Majorana neutrino mass $M_{R}$.
The inverse of $M_{R}$ is now given by
\begin{equation}
M_{R}^{-1}=\left(\begin{array}{ccc} (\epsilon_{1}\epsilon_{3}-\mu^{2})&(\mu\rho-\epsilon_{3}\sigma)& (\sigma\mu-\epsilon_{2}\rho\\
(\mu\rho-\epsilon_{3}\sigma) & (\epsilon_{1}\epsilon_{3}-\rho^{2}) & (\sigma\rho-\epsilon_{1}\mu)\\
(\sigma\mu-\epsilon_{2}\rho)& (\sigma\rho-\epsilon_{1}\mu)& (\epsilon_{1}\epsilon_{2}-\sigma^{2})
\end{array}\right) (Dv_{R})^{-1}
\end{equation}
where
\begin{equation}
D = det|M_{R}| = ( \epsilon_{1}\epsilon_{2}\epsilon_{3} + 2 \rho\sigma\mu - \epsilon_{1}\mu^{2}
- \epsilon_{2}\rho^{2}-\epsilon_{3}\sigma^{2}).\nonumber
\end{equation}
For simplicity, we maintain the hierarchical structure in  $M_{R}$  by  making  the  substitution
$\epsilon_{2}=\epsilon^{2}, \mu= b\epsilon , \epsilon_{3} = 1 , \epsilon_{1}< \epsilon_{2}$ in Eqs. (6) and (7):
\begin{equation}
M_{R}=
\left(\begin{array}{ccc}
\epsilon_{1} & \sigma & \rho\\
\sigma & \epsilon^{2}& b\epsilon\\
\rho & b\epsilon & 1
\end{array}\right) v_{R},
\end{equation}
\begin{equation}
M_{R}^{-1}\sim
\left(\begin{array}{ccc}
\frac{\epsilon^{2}}{\rho^{2}}(1-b^{2}) &-\frac{ 1}{\rho^{2}}(\sigma-\rho b\epsilon) &-\frac{ 1}{\rho^{2}}(\rho\epsilon^{2}-\sigma b \epsilon)\\
-\frac{1}{\rho^{2}}(\sigma-\rho b \epsilon) &-(1-\epsilon_{1}/\rho^{2}) &\frac{ \sigma}{\rho}(1-b\epsilon\epsilon_{1}/\rho\sigma)\\
-\frac{1}{\rho^{2}}(\rho\epsilon^{2}-\sigma b \epsilon) &\frac{ \sigma}{\rho}(1-b\epsilon\epsilon_{1}/\rho\sigma) &-\frac{ \sigma^{2}}{\rho^{2}}(1-\epsilon_{1} \epsilon^{2}/\sigma^{2})
\end{array}\right)
\end{equation}
where the denominator $(Dv_{R})$ has been supressed here as we are interested 
in the form of texture only.
The Dirac neutrino mass matrix $m_{LR}$ appeared in the see-saw formula (1),
can be either the charged lepton mass matrix $m_{e}$ (referred to as case (i)) 
or the up-quark mass matrix $m_{u}$ (referred  to as case (ii))
depending on the particular model of the SO(10) GUT and the content of the 
Higgs fields [16,17] employed.
\\

\underline{Case (i) where $m_{LR}= m_{e}$}

 Since the charged lepton masses
 have the ratio $m_{\tau}: m_{\mu}:m_{e}=1:\lambda^{2}:\lambda^{6}$ [13,18], our choice for the diagonal $m_{LR}$ is
\begin{equation}
m_{LR}=
\left(\begin{array}{ccc}
\lambda^{6} & 0 & 0 \\
0 & \lambda^{2} & 0 \\
0 & 0 &  1
\end{array}\right) m_{\tau}
\end{equation}
where we take the value of the Wolfenstein parameter[19] 
$\lambda\approx 0.22$.
Substituting Eqs.(10) and (11) to (1), we have
\begin{equation}
m_{LL}\sim \left(\begin{array}{ccc}
-\frac{\epsilon^{2}}{\rho^{2}} (1-b^{2})\lambda^{8} & \frac{1}{\rho^{2}} (\sigma-\rho b\epsilon)\lambda^{4}
& \frac{1}{\rho^{2}} (\rho\epsilon^{2}-\sigma b\epsilon)\lambda^{2}\\
\frac{1}{\rho^{2}} (\sigma-\rho b\epsilon)\lambda^{4} & (1-\frac{\epsilon_{1}}{\rho^{2}})
& -\frac{\sigma}{\rho}
(1-\frac{\epsilon_{1} b\epsilon}{\rho\sigma}) \lambda^{-2}\\
\frac{1}{\rho^{2}}(\rho\epsilon^{2}-\sigma b\epsilon) \lambda^{2}
& -\frac{\sigma}{\rho}(1-\frac{\epsilon_{1} b\epsilon}{\rho\sigma})\lambda^{-2}
& \frac{\sigma^{2}}{\rho^{2}}(1-\frac{\epsilon_{1}\epsilon^{2}}{\sigma^{2}}) \lambda^{-4}
\end{array}\right) m_{0}
\end{equation}
where $m_{0}$ is the overall proportionality constant.
\\

\underline{Case (ii) where $m_{LR}=m_{u}$}

The up-quark masses have the mass ratio  $ m_{t}: m_{c}: m_{u}= 1:\lambda^{4}:\lambda^{8}$ [13,18] and $m_{LR}$ has the form
\begin{equation}
m_{LR}=
\left(\begin{array}{ccc}
\lambda^{8} & 0 & 0 \\
0 & \lambda^{4} & 0 \\
0 & 0 & 1
\end{array}\right) m_{t}
\end{equation}
and the corresponding left handed Majorana neutrino mass matrix can be 
calculated as
\begin{equation}
m_{LL}\sim \left(\begin{array}{ccc}
-\frac{\epsilon^{2}}{\rho^{2}} (1-b^{2})\lambda^{8} & \frac{1}{\rho^{2}} (\sigma-\rho b\epsilon)\lambda^{4}
& \frac{1}{\rho^{2}} (\rho\epsilon^{2}-\sigma b\epsilon)\\
\frac{1}{\rho^{2}} (\sigma-\rho b\epsilon)\lambda^{4} & (1-\frac{\epsilon_{1}}{\rho^{2}})
& -\frac{\sigma}{\rho}
(1-\frac{\epsilon_{1} b\epsilon}{\rho\sigma}) \lambda^{-4}\\
\frac{1}{\rho^{2}}(\rho\epsilon^{2}-\sigma b\epsilon)
& -\frac{\sigma}{\rho}(1-\frac{\epsilon_{1} b\epsilon}{\rho\sigma})\lambda^{-4}
& \frac{\sigma^{2}}{\rho^{2}}(1-\frac{\epsilon_{1}\epsilon^{2}}{\sigma^{2}}) \lambda^{-8}
\end{array}\right) m_{0}
\end{equation}
In general we can have a hierarchical $m_{LR}$ in the diagonal form 
and a corresponding $m_{LL}$ can be calculated. $m_{LL}$
in Eqs.(12) and (14) for case (i) and case (ii) can be reduced to a 
form which can generate bimaximal mixings suitable for the explanation 
for LMA solution and
 atmospheric neutrino oscillation. This can be done in both hierarchical and 
inverted hierarchical schemes of neutrino mass  patterns.
\section{(A) Hierarchical neutrino mass model}

In the hierarchical model we have the neutrino mass pattern $m_{\nu3}> m_{nu2}> m_{\nu1}$ and the splitting parameter
$\xi= \Delta m_{12}^{2}/\Delta m_{23}^{2}\simeq m_{\nu2}^{2}/m_{\nu3}^{2}$. We analyse $m_{LL}$ for case (i) and case (ii) given
in Eqs.(12) and (14) respectively for generating bimaximal mixings.
\\

\underline{Case (i) where  $m_{LR}= m_{e}$:}

In Eq.(12) we make suitable  choices for the parameters in $m_{LL}$ as follows:

(a)$ \rho=\lambda^{5}, \sigma= \lambda^{7},\epsilon= \lambda^{3}, \epsilon_{1}= \lambda^{11}$

\begin{equation}
m_{LL}\sim
\left(\begin{array}{ccc}
-\lambda^{4}(1-b^{2})  &  \lambda(1-b\lambda)  &   \lambda^{3}(1-b\lambda^{-1})\\
 \lambda(1-b\lambda)   &  (1-\lambda)          &   -(1-b\lambda^{2})\\
\lambda^{3}(1-b\lambda^{-1})  &  -(1-b\lambda^{2}) & (1-\lambda^{3})
\end{array}\right)m_{0}
\end{equation}

This simplifies to, for b=0,

\begin{equation}
m_{LL}\sim
\left(\begin{array}{ccc}
-\lambda^{4}  &  \lambda  &  \lambda^{3}\\
 \lambda      &  (1-\lambda) & -1 \\
\lambda^{3}    & -1          & (1-\lambda^{3})
\end{array}\right)m_{0}
\end{equation}
which predicts $\xi= \Delta m_{12}^{2}/\Delta m_{23}^{2}\simeq 0.0134, sin^{2}2\theta_{12}=0.8987,  sin^{2}2\theta_{23}=0.9809, |V_{e3}|=0.074$. 
The corresponding texture of  $M_{R}$ is

\begin{equation}
M_{R}=
\left(\begin{array}{ccc}
\lambda^{11}  &  \lambda^{7}  &   \lambda^{5}\\
 \lambda^{7}   &  \lambda^{6}  &   0\\
\lambda^{5}  &  0  &  1
\end{array}\right)v_{R}
\end{equation}
(b) The choice  $ \rho=\lambda^{4}, \sigma= \lambda^{6},\epsilon= \lambda^{3}, \epsilon_{1}=- \lambda^{10}, b=\frac{1}{\lambda^{2}}$
gives

\begin{equation}
m_{LL}\sim
\left(\begin{array}{ccc}
\lambda^{2}-\lambda^{6}  &  -\lambda+\lambda^{2}  &   -\lambda+\lambda^{4}\\
 -\lambda+\lambda^{2}  &   1+\lambda^{2}          &   -(1+\lambda)\\
 -\lambda+\lambda^{4}  &    -(1+\lambda)          &    (1+\lambda^{4})
\end{array}\right)m_{0}
\end{equation}
This predicts    $\xi= \Delta m_{12}^{2}/\Delta m_{23}^{2}=0.018, sin^{2}2\theta_{12}=0.8372, sin^{2}2\theta_{23}=0.9994, |V_{e3}|=0.014$.
The corresponding $M_{R}$ is

\begin{equation}
M_{R}=
\left(\begin{array}{ccc}
-\lambda^{10}  &  \lambda^{6}  &   \lambda^{4}\\
 \lambda^{6}   &  \lambda^{6}  &   \lambda\\
\lambda^{4}  &  \lambda  &  1
\end{array}\right)v_{R}
\end{equation}

\underline{ Case (ii) where $ m_{LR}= m_{u}$}

We again examine the texture of $m_{LL}$ in Eq.(14) with the following 
choices of parameters.

(c) The choice $ \rho=\lambda^{7}, \sigma= \lambda^{11},\epsilon=
 \lambda^{5}, M_{1}= \lambda^{15}$ leads to the texture of $m_{LL}$ in Eq.(15),
and for b=0, it again reproduces the same $m_{LL}$ given in Eq.(16), and the 
 corresponding $M_{R}$ is

\begin{equation}
M_{R}=
\left(\begin{array}{ccc}
\lambda^{15}  &  \lambda^{11}  &   \lambda^{7}\\
 \lambda^{11}   &  \lambda^{10}  &        0 \\
\lambda^{7}  &   0  &  1
\end{array}\right)v_{R}
\end{equation}

(d) The choice  $ \rho=\lambda^{7}, \sigma= \lambda^{11},\epsilon= \lambda^{4}, \epsilon_{1}=- \lambda^{15}, b=\lambda$
gives

\begin{equation}
m_{LL}\sim
\left(\begin{array}{ccc}
-\lambda^{2}(1-\lambda^{2})  &  \lambda(1-\lambda)  &   \lambda(1-\lambda)\\
 \lambda(1-\lambda)  &   (1-\lambda)          &   -(1+\lambda^{2})\\
 \lambda(1-\lambda)  &    -(1+\lambda^{2})          &    (1+\lambda)
\end{array}\right)m_{0}
\end{equation}
which yields     $\xi= \Delta m_{12}^{2}/\Delta m_{23}^{2}=0.01326, sin^{2}2\theta_{12}=0.8326, sin^{2}2\theta_{23}=0.99, |V_{e3}|\approx0.0$.
The corresponding $M_{R}$ is

\begin{equation}
M_{R}=
\left(\begin{array}{ccc}
-\lambda^{15}  &  \lambda^{11}  &   \lambda^{7}\\
 \lambda^{11}   &  \lambda^{8}  &   \lambda^{3}\\
\lambda^{7}  &  \lambda^{3}  &  1
\end{array}\right)v_{R}
\end{equation}

\section{(B) Inverted hierarchical mass model}

The left-handed Majorana neutrino mass matrix $m_{LL}$ in Eqs.(12) and (14)
 for case(i) and case(ii) respectively, can also be expressed in
terms of mass matrix which predicts inverted hierarchical mass pattern  $m_{\nu1}\geq m_{\nu2}\gg m_{\nu3}$. Thus it has the texture[20,21]

 \begin{equation}
 m_{LL}=
 \left(\begin{array}{ccc}
 \delta_{1}  &  a  &   c\\
  a   &  \delta_{2}  &   \delta_{2}\\
 c  &  \delta_{2}  &  \delta_{2}
 \end{array}\right)m_{0},
 \end{equation}
 where $\delta_{1}, \delta_{2}\ll 1; a, c \approx 1$
 having the eigenvalues $m_{1}$, $m_{2}$, $m_{3}$ such that   
$m_{1}\approx m_{2}\approx m_{0}, \Delta m_{12}^{2}\approx
m_{0}^{2}(\delta_{2} + \delta_{1}), m_{3}\approx m_{0} \delta_{2} , 
\Delta m_{23}^{2} \approx m_{0}^{2}$.
This gives the splitting parameter,
$\xi = \Delta m_{12}^{2}/\Delta m_{23}^{2}\approx 2(m_{1} - m_{2})/m_{2}
\approx (\delta_{1}+ \delta_{2})$,
   and mixing angles are $ tan\theta_{23}\approx a/c \approx 1, tan\theta_{12}\approx1, \theta_{13}\simeq \delta_{2}$.
\\

\underline{Case (i) where  $m_{LR}=m_{e}$:}

We make particular choices of the parameters in Eq.(12) for  obtaining 
the form of the texture of $m_{LL}$ given in Eq.(23)
for inverted hierarchical model.

(e) The choice  $ \rho=\lambda^{9}, \sigma= \lambda^{11},\epsilon= \sigma/\rho, \epsilon_{1}=- \rho^{2}, b=-1$
gives

\begin{equation}
m_{LL}=
\left(\begin{array}{ccc}
 0  &  1  &   1\\
 1   &  \lambda^{3}  &   0\\
 1  &  0 &  \lambda^{3}
 \end{array}\right)m_{0},
 \end{equation}
 giving    $\xi= \Delta m_{12}^{2}/\Delta m_{23}^{2}=0.014, sin^{2}2\theta_{12}=0.9999, sin^{2}2\theta_{23}=1.00, |V_{e3}|\approx0.0$. The corresponding $M_{R}$ is given by 

  \begin{equation}
  M_{R}=
  \left(\begin{array}{ccc}
  -\lambda^{18}  &  \lambda^{11}  &   \lambda^{9}\\
   \lambda^{11}   &  \lambda^{4}  &   -\lambda^{2}\\
  \lambda^{9}  &  -\lambda^{2}  &  1
  \end{array}\right)v_{R}
  \end{equation}

 (f) The choice  $ \rho=\lambda^{10}, \sigma= \lambda^{12},\epsilon= \sigma/\rho, \epsilon_{1}=\rho^{2}/\lambda, b=-1$
 leads to

 \begin{equation}
 m_{LL}=
 \left(\begin{array}{ccc}
  0  &  1  &   1\\
  1   &  -(\lambda^{3}-\lambda^{4})/2  &  -(\lambda^{3}+\lambda^{4})/2\\
  1  & -(\lambda^{3}+\lambda^{4})/2  &  -(\lambda^{3}-\lambda^{4})/2
  \end{array}\right)m_{0},
  \end{equation}
giving    $\xi= \Delta m_{12}^{2}/\Delta m_{23}^{2}=0.0151, sin^{2}2\theta_{12}=0.9999, sin^{2}2\theta_{23}=1.00, |V_{e3}|\approx0.0$. The corresponding $M_{R}$ is

  \begin{equation}
  M_{R}=
  \left(\begin{array}{ccc}
  \lambda^{19}  &  \lambda^{12}  &   \lambda^{10}\\
   \lambda^{12}   &  \lambda^{4}  &   -\lambda^{2}\\
  \lambda^{10}  &  -\lambda^{2}  &  1
  \end{array}\right)v_{R}
  \end{equation}

 (g) The choice  $ \rho=\lambda^{10}, \sigma= \lambda^{12},\epsilon= \sigma/\rho, \epsilon_{1}= 0, b=-(1+\lambda^{7})$
 leads to

 \begin{equation}
 m_{LL}=
 \left(\begin{array}{ccc}
  \lambda^3  &  1  &   1\\
  1   &  \lambda^{4}/2  &   -\lambda^{4}/2\\
  1  & -\lambda^{4}/2  &  \lambda^{4}/2
  \end{array}\right)m_{0},
  \end{equation}
giving    $\xi= \Delta m_{12}^{2}/\Delta m_{23}^{2}=0.0151, sin^{2}2\theta_{12}=0.9999, sin^{2}2\theta_{23}=1.00, |V_{e3}|\approx0.0$. The corresponding $M_{R}$ is

   \begin{equation}
   M_{R}=
   \left(\begin{array}{ccc}
   0  &  \lambda^{12}  &   \lambda^{10}\\
  \lambda^{12}   &  \lambda^{4}  &   -(\lambda^{2}+\lambda^{9})\\
  \lambda^{10}  &  -(\lambda^{2}+\lambda^{9})  &  1
  \end{array}\right)v_{R}
  \end{equation}

\underline{Case (ii) where $m_{LR}= m_{u}$}

$m_{LL}$ in Eq.(14) is again expressed in the form given in Eq.(23) with 
the following choices of the parameters.

(h) The choice  $ \rho=\lambda^{11}, \sigma= \lambda^{15},\epsilon= \sigma/\rho, \epsilon_{1}=-\rho^{2} , b=-1$
 leads to the same $m_{LL}$ given in Eq.(24), and the corresponding $M_{R}$ is

\begin{equation}
  M_{R}=
  \left(\begin{array}{ccc}
  -\lambda^{22}  &  \lambda^{15}  &   \lambda^{11}\\
   \lambda^{15}   &  \lambda^{8}  &   -\lambda^{4}\\
  \lambda^{11}  &  -\lambda^{4}  &  1
  \end{array}\right)v_{R}
  \end{equation}

(i) The choice  $ \rho=\lambda^{12}, \sigma= \lambda^{16},\epsilon= \sigma/\rho, \epsilon_{1}=\rho^{2}/\lambda , b=-1$
 leads to the same $m_{LL}$ given in Eq.(26), and the corresponding $M_{R}$ is

\begin{equation}
  M_{R}=
  \left(\begin{array}{ccc}
  \lambda^{23}  &  \lambda^{16}  &   \lambda^{12}\\
   \lambda^{16}   &  \lambda^{8}  &   -\lambda^{4}\\
  \lambda^{12}  &  -\lambda^{4}  &  1
  \end{array}\right)v_{R}
  \end{equation}

(j) The choice  $ \rho=\lambda^{12}, \sigma= \lambda^{16},\epsilon= \sigma/\rho, \epsilon_{1}=0, b=-(1+\lambda^{7}) $
 leads to the same $m_{LL}$ given in Eq.(28), and the corresponding $M_{R}$ is

\begin{equation}
  M_{R}=
  \left(\begin{array}{ccc}
  0  &  \lambda^{16}  &   \lambda^{12}\\
   \lambda^{16}   &  \lambda^{8}  &   -(\lambda^{4}+\lambda^{12})\\
  \lambda^{12}  &  -(\lambda^{4}+\lambda^{12})  &  1
  \end{array}\right)v_{R}
  \end{equation}

\section{Summary and discussion}
The bimaximal mixings which explain both LMA MSW solution of the solar
 neutrino 
anomaly and the atmospheric neutrino solution, have been
generated through the see-saw mechanism using the non-zero texture of the 
right-handed Majorana neutrino mass matrix $M_{R}$ and
the diagonal form of the Dirac neutrino mass matrix. The left-handed 
Majorana mass matrix $m_{LL}$ has been derived for both hierarchical 
and inverted hierarchical models
of neutrino mass patterns using the proper choices of the values of the 
parameters in $M_{R}$. In our analysis we have considered both
possibilities where the Dirac neutrino mass matrix $m_{LR}$ is either the 
charged-lepton or the up-quark mass matrix. For illustration, we present
a few representative examples for each case.

In case of hierarchical model all the predictions are excellent with the values
$\xi=0.014, sin^{2}2\theta_{12}=0.83,  sin^{2}2\theta_{23}=0.99$. In case of inverted hierarchical model, the predictions
are $\xi=0.015, sin^{2}2\theta_{12}=0.9999, sin^{2}2\theta_{23}=1.000, |V_{e3}|=0.0$ in  almost all examples.
But the LMA MSW solution [5,6,7] shows that $sin^{2}\theta_{12}$ is large but not maximal, and has a upper bound 0.98. There
is indeed an inherent problem in the inverted hierarchical model [21], and such problem can be remedied partly with the
choice of the particular form of the charged lepton mass matrix  $m_{e}$ which can lower the $\theta_{12}$ mixing angle.
It has also been observed that under renormalisation group analysis in MSSM,
 $ sin^{2}2\theta_{12}$ in case of inverted hierarchical model,
increases with energy. All these give enough scope to accomodate the present result for the explanation of LMA MSW solution
at low energies[22]. The present work exploits the origin of the bimaximal mixings from the non-zero texture of $M_{R}$
in a model independant way, whereas in the earlier attempts [6,11-14], the  non-diagonal forms of both $m_{LR}$ and $M_{R}$,
or at least $m_{LR}$, have been employed for generating bimaximal mixings. In this paper only a few representative
examples are presented. Many more examples for other choices of parameters are also possible for generating bimaximal mixings.
 The present analysis may be a useful guide for building models under the framework of grand unified theories with extended U(1)
 symmetry.


\begin{thebibliography}{99}
\bibitem{1} The Super-Kamiokande Collaboration, Phys. Rev. Lett.{\bf 86},5656(2001).
\bibitem{2} H.Sobel, Talk presented at the XIX International Conference on Neutrino Physics and Astrophysics, Sudbury,
Canada,June 16-21,2000.
\bibitem{3} S.F.King, {\bf hep-ph/0105261} and further references therein.
\bibitem{4} J.N.Bahcall, P.I.Krastev and A.Yu.Smirnov, JHEP {\bf 0105},015 (2001).
\bibitem{5} J.N.Bahcall, M.C.Gonzalez-Garcia, C.Pena-Garay, {\bf hep-ph/0106258}.
\bibitem{6} Ilja Dorsner and S.M.Barr, {\bf hep-ph/0108168}; S.M.Barr and Ilja Dorsner, Nucl.Phys.{\bf B585},79(2000), {\bf hep-ph/0003058}.
\bibitem{7} M.Apollonio et al.,Phys-Lett.{\bf B466},415(1999).
\bibitem{8} LSND collaboration, Phys.Rev.Lett.{\bf 77},3082(1996);{\it ibid}{\bf 81},1774(1998).
\bibitem{9} M.Gell-Mann, P.Ramond, R.Slansky, in : Supergravity, North-Holland, Amsterdam,1979; 
 T.Yanagida, in : Proc.of the workshop on unified Theory and Baryon Number of the Universe, KEK, Japan,1979;  R.N. Mohapatra, G. Senjanovic, Phys.Rev.Lett. {\bf 44},912 (1980).
\bibitem{10} Mu-Chun Chen and K.T.Mahanthappa, {\bf hep-ph/0005292}.
\bibitem{11}  S.F.King and M.Oliveira, {\bf hep-ph/0009287}.
\bibitem{12} B.R.Desai, U.Sarkar, A.R.Vaucher, {\bf hep-ph/0007346}.
\bibitem{13} G.Altarelli and F.Feruglio, Phys.Rept.{\bf 320},295 (1999), {\bf hep-ph/9905536},
G.Altarelli,F.Feruglio and I.Masina, Phys.Lett.{\bf B472}, 382 (2000), {\bf hep-ph/9907532}.
\bibitem{14} Z.Maki, M. Nakagawa and S.Sakata, Prog.Theor.Phys.{\bf28},870(1972).
\bibitem{15} S.F.King and N.Nimai Singh, Nucl.Phys.{\bf B591}, 3 (2000).
\bibitem{16} K.S.Babu, B.Dutta, R.N.Mahapatra, Phys.Lett.{\bf B458}, 93 (1999),
{\bf hep-ph/9904366}, E.Malkawi,Phys.Rev.{\bf D61},013006(1999).
\bibitem{17} K.S.Babu and S.M.Barr,{\bf hep-ph/0004118}.
\bibitem{18} P.M.Fishbane and P.Kaur, J.Phys.G: Nucl.Part.Phys.{\bf 25},1629(1999).
\bibitem{19} L.Wolfenstein, Phys.Rev.Lett.{\bf 51},1945(1983).
\bibitem{20} R.Barbieri, L.Hall, D.Smith, A.Strumia and N.Weiner, 
 JHEP {\bf 9812},17 (1998), 
{\bf hep-ph/9807235}; A.S.Joshipura, S.D.Rindani, Eur.Phys.J. {\bf C14},85(2000);
 R.N.Mahapatra, A.Perez-Lorenzana, C.A.de.S.Pires,  Phys.Lett. {\bf B474}, 355 (2000), {\bf hep-ph/9911395};
 Q.Shafi and Z.Tavartkiladze, Phys.Lett. {\bf B482},145(2000), {\bf hep-ph/0002150}.
\bibitem{21} S.F.King and N.Nimai Singh, Nucl.Phys.{\bf B596},81(2001).
\bibitem{22} Mahadev Patgiri and N.Nimai Singh, in preparation (2001).
\end{thebibliography}
\end{document}